# Atomistic Modeling of Functionalized Magnetite Surfaces with Oxidation States

Emre Gürsoy, Robert H. Meißner, and Gregor B. Vonbun-Feldbauer*

ACCESS | Metrics & More | Article Recommendations | Supporting Information

**ABSTRACT:** Understanding the atomic structure of magnetite-carboxylic acid interfaces is crucial for tailoring nanocomposites involving this interface. We present a Monte Carlo (MC)-based method utilizing iron oxidation state exchange to model magnetite interfaces with tens of thousands of atoms, scales typically inaccessible by electronic structure calculations. Charge neutrality is ensured by the oxidation of Fe ions. The MC approach allows magnetite to adapt to its environment at interfaces without requiring interface-specific rescaling of force-field parameters. This enables a simple, versatile method. By comparing adsorption sites, layer distances, and bond lengths with results from electronic structure calculations and experiments, we validated the accuracy of our method. We found that the oxidation state distribution and, consequently, binding site preference depend on coverage and surface thickness, with a critical thickness signaling the transition from layered to bulk-like oxidation states. The method ensures seamless compatibility with popular biomolecular force fields providing transferability and simplifying the study of magnetite interfaces in general.

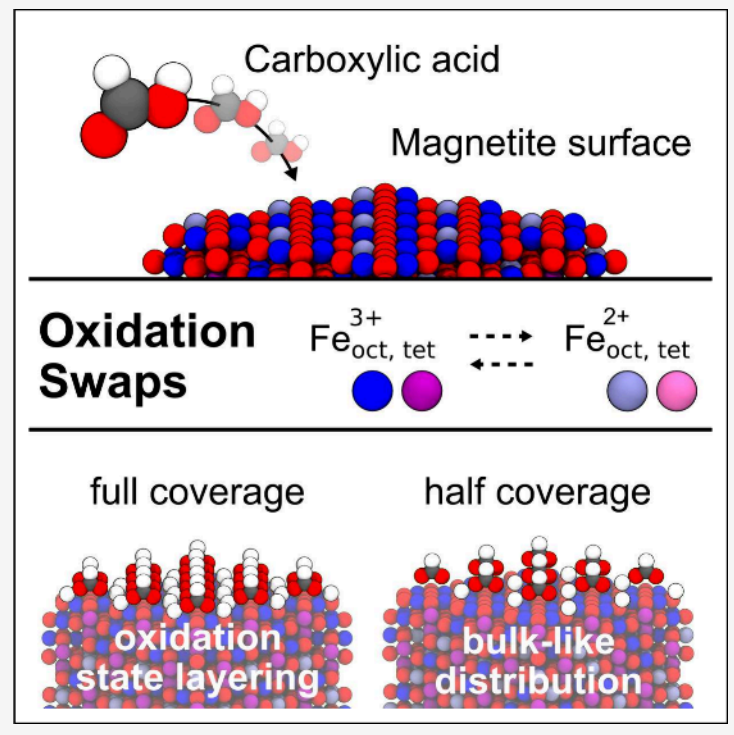
Carboxylic acid
Magnetite surface
Oxidation Swaps
$Fe^{3+}_{oct, tet}$
$Fe^{2+}_{oct, tet}$
full coverage
half coverage
oxidation state layering
bulk-like distribution

Magnetite is a biocompatible material that finds various applications spanning from drug delivery[1] and therapeutic agents[2] to magnetic resonance imaging (MRI).[3] Magnetite interfaces play a crucial role in building a cleaner and more sustainable future, including pesticide removal from water,[4] Fischer−Tropsch synthesis,[5] and the water-gas shift reaction.[6] Moreover, functionalized magnetite nanoparticles, e.g. with oleic acid, serve as fundamental building blocks for hierarchical nanocomposites.[7,8] Those nanocomposites have exceptional mechanical properties[9] that can be fine-tuned by modifying nanoparticle morphology,[10,11] introducing additional hierarchical levels,[12] or altering ligand reactivity.[13]

In general, magnetite exhibits two dominant surface facets, the (001) and (111) surfaces, which are often observed at magnetite nanoparticles due to their low surface energies.[10] The stability and morphology of the (111) surface strongly depends on the surrounding environment and the specific preparation conditions.[14−19] For the (001) surface two surface models are commonly used, the distorted bulk truncation (DBT)[20] and the subsurface cation vacancy (SCV) reconstruction.[21] While the SCV reconstruction is usually found under ultrahigh-vacuum conditions, the DBT can be stabilized in the presence of adsorbates like hydrogen and formic acid.[22−25] On (001) magnetite surfaces under UHV conditions, formic acid (HCOOH) undergoes dissociative adsorption and dissociates into a formate ($HCOO^{-}$) and a proton ($H^{+}$), where the formate prefers a bidentate binding mode. On the (001)-DBT surface, two distinct adsorption site are observed: formate adsorbs either next to an $Fe_{tet}$ ion, the "tet" adsorption site, or in a region between two $Fe_{tet}$ ions, the "int" adsorption site (cf. Figure 1).[23,24,26] To harness the potential of magnetite based nanocomposites, intensive efforts were made to shed light on magnetite interfaces,[27] particularly those involving formic acids serving as the archetype for carboxylic ligands.[18,23,24,26] Although this work focuses on the (001)-DBT surface as a model system due to the availability of high-quality experimental and computational data, the method described herein is considered to be applicable to any magnetite-organic adsorbate interface.

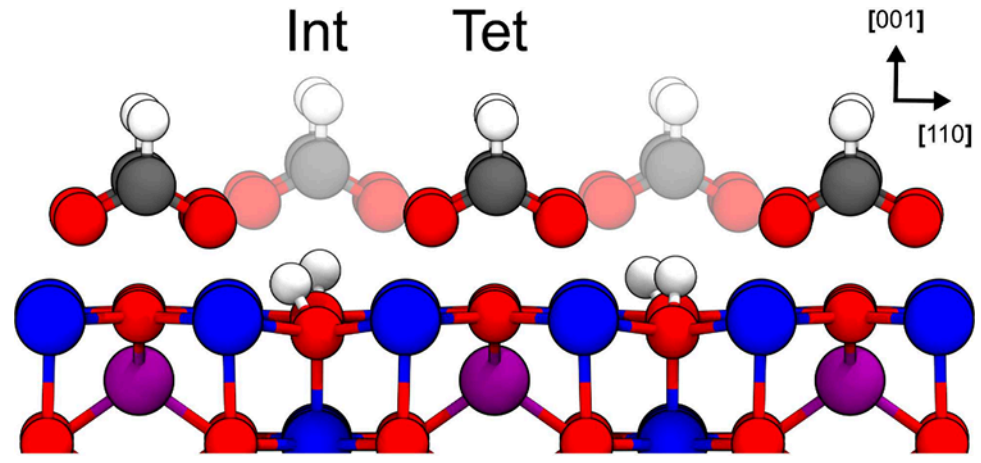

**Figure 1.** Formic acid adsorption sites on (001) magnetite surface. Adsorption sites, "tet" and "int", are highlighted. Color code: $Fe^{3+}_{oct}$, dark blue; $Fe^{3+}_{tet}$, purple; O, red; H, white; C, gray.

In atomistic models of magnetite interfaces, atoms are often described by point charges, whereby the parameters describing these point charges are tailored either to specific interfaces, such as carboxylic acids,[26,28] water,[29] and phosphonic acids,[4] or to specific morphologies, such as nanoparticles.[30,31] The parametrization of force fields is typically done on the basis of electronic structure calculations or experiments.

The dissociation of carboxylic acids upon adsorption onto magnetite requires a specific parametrization of the proton transfer. Common drawbacks of many force fields are incorrect energies and the frequent appearance of excess charges in the case of chemical reactions.[32,33] The latter is often compensated for by adding or removing charge carriers within the system to maintain charge neutrality. This can be achieved, e.g., by adding ions,[34] dissociating water on the surface to form hydroxyl groups,[35] distributing the excess charge on some atoms,[26] or creating deprotonated charged surface sites.[32] While adding or removing charge carriers secures the charge neutrality, it changes the overall composition of the system.

One of the main goals of this work is to present a versatile and simple approach, which needs as little as possible interface or surface specific reparametrization of standard force field parameters, here taken from GAFF[36,37] and ClayFF[38] and shown in more detail in the Supporting Information (SI). Using standard GAFF and ClayFF partial charges for formic acid and formate, and magnetite, respectively (cf. Konuk et al.[26]), results in a negatively charged system: formate contributes −1.0 $e$ to the overall charge, hydroxyl hydrogen $H_O$ contributes +0.425 $e$, and the bridging oxygen (O) in magnetite turns into a hydroxyl oxygen ($O_H$), changing its charge by +0.05 $e$, where $e$ represents the elementary charge. This distribution with essential contributions on the formate and $H_O$ and only a minor contribution on $O_H$ agrees qualitatively with Bader charge analysis on DFT data in our previous work.[26] Summing up the contributions for the dissociative adsorption of a single formic acid molecule results in an excess charge of −0.525 $e$ (cf. $\Delta q$ in Figure 2b). Previously the necessary compensation charge was distributed among the atoms in proximity of the reaction site based on reference density functional theory (DFT) calculations.[26,27] This makes the parametrization interface specific, meaning it has to be redone for each surface facet and termination, and limits its transferability. In contrast to our previous approach in ref 26 and similarly in recent literature,[28] we use here standard ClayFF charges for magnetite as far as possible (see SI for more details) and do not rely on modified partial charges which need to be specifically derived for each surface, like (001)-DBT, (001)-SCV, or (111).

We present a simple but effective atomistic simulation method that neither requires additional charge carriers nor interface specific magnetite parameters which, e.g., depend on the specific surface.[26] This method is based on three steps: (1) including the charge transfer from adsorbates to magnetite by oxidizing $Fe^{2+}$ ions (cf. Figure 2a,b), (2) minimizing the potential energy of the adsorbates, and (3) using Monte Carlo (MC) to swap the oxidation states of the Fe ions. Steps 2 and 3 are carried out repeatedly until convergence (cf., Figure 2d). This process is referred to as "oxidation state minimization", as it entails forcing the oxidation states of Fe ions to adapt to the adsorbate geometry. Details of oxidation state minimization are given in the SI. This represents an extension of our previous method[39] which was developed for investigating bulk magnetite, and unfunctionalized magnetite surfaces and nanoparticles.

More specifically, the excess charge in step 1 from the dissociation of a formic acid on its respective adsorption site is compensated by oxidizing $Fe^{2+} \rightarrow Fe^{3+}$, where $q^{3+} - q^{2+} =$ 0.525 $e$. This oxidation state change ensures charge neutrality (cf. $\Delta q$ in Figure 2c) of the overall system when formic acid dissociates and a hydroxyl is formed on the surface. Note that in general, we assume that the missing $Fe^{2+}$ on the octahedral places due to surfaces and adsorption are compensated and, e.g., found in defects in the bulk of real systems. This assumption may not hold for small systems where the $Fe^{2+}$/$Fe^{3+}$ ratio is far from ideal stoichiometry. Force field parameters describing magnetite, formate, and surface hydroxides are presented in Table SI1. For comparison, DFT calculations were performed on functionalized magnetite slabs. Compared to our previous formic acid adsorption studies,[24,26] we increased the system size and removed symmetry constraints following our work on a bare magnetite surface.[39] Details on the DFT calculations can be found in the SI.

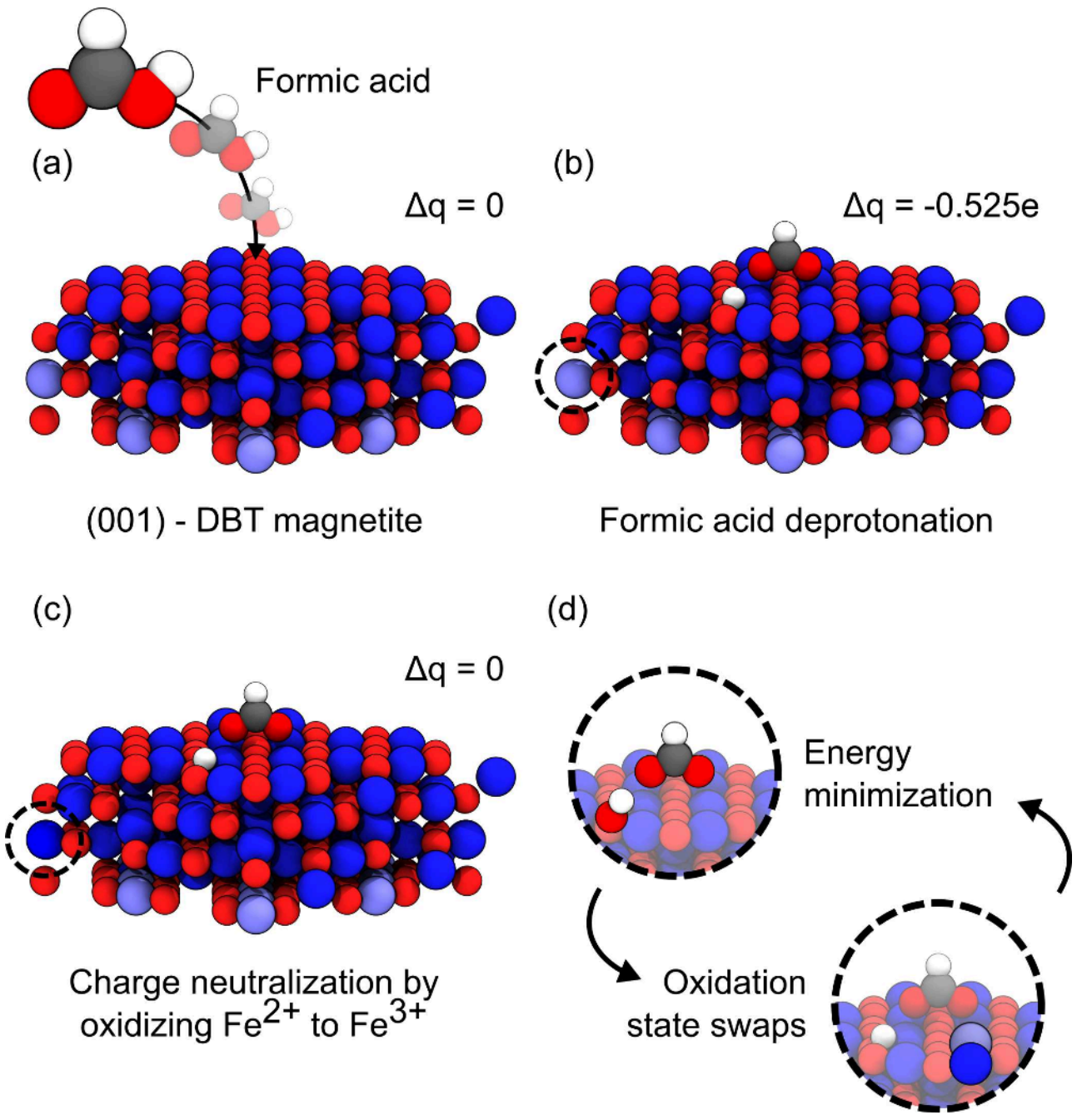

**Figure 2.** Formic acid adsorption and oxidation state minimization at magnetite interface. (a) Schematic representation of formic acid adsorption on magnetite surface. The overall system is charge neutral. (b) Formic acid deprotonates to formate and a hydrogen. As the result, the total charge of the system (adsorbate + magnetite) becomes slightly negative (−0.525 $e$). (c) By oxidizing an $Fe^{2+}$ to $Fe^{3+}$ the system becomes charge neutral again. (d) Through a cyclic process of oxidation state swaps and energy minimization, the minimized oxidation state configurations are obtained. Color code: $Fe^{3+}_{oct}$, dark blue; $Fe^{2+}_{oct}$, light blue; O, red; H, white; C, gray.

Oxidation state minimization was applied on functionalized magnetite slabs with thickness ranging from 9L (∼1 nm) to 65L (∼7 nm), where L stands for iron layers. Formate molecules are located at "tet" and "int" binding sites for both half and full formate coverage. Full coverage is defined here as every surface Fe atom having a bond to a formate O atom. Some structures are shown as examples in Figure 3. It is important to note that both the top and bottom magnetite surfaces of the slab used in the simulations are functionalized. Due to the long-range electrostatics, functionalizing only one surface results in an unwanted dipole and in an unrealistic asymmetric oxidation state distribution (cf. Figure SI1). At room temperature, all $Fe_{tet}$ in magnetite are usually $Fe^{3+}$. Hence, all $Fe_{tet}$ are fixed to $Fe^{3+}$ and excluded from the oxidation swaps. Consequently, we focus only on $Fe_{oct}$, which can be both $Fe^{3+}$ or $Fe^{2+}$. At elevated temperatures this might not be a reasonable approach and swaps should also be done for $Fe_{tet}$.[40] To assess whether or not a given layer exhibits a bulk-like oxidation state distribution with $n_{Fe_{oct}{}^{3+}}/n_{Fe_{oct}} = 0.5$, the $Fe^{3+}$ ratio within each $Fe_{oct}$ layer was calculated and is shown in Figure 3. Since in the DFT calculations the OH molecules at the top and bottom interfaces face in opposite directions, following the point symmetry of the magnetite slab, we focus on these for consistency. Nevertheless, we have checked in Figures SI2 and SI3 if the direction of the surface hydroxyls has an effect on our simulations. No significant differences were found. Atomistic structures of both same-directional and opposite-directional OH models can be seen in Figure SI4.

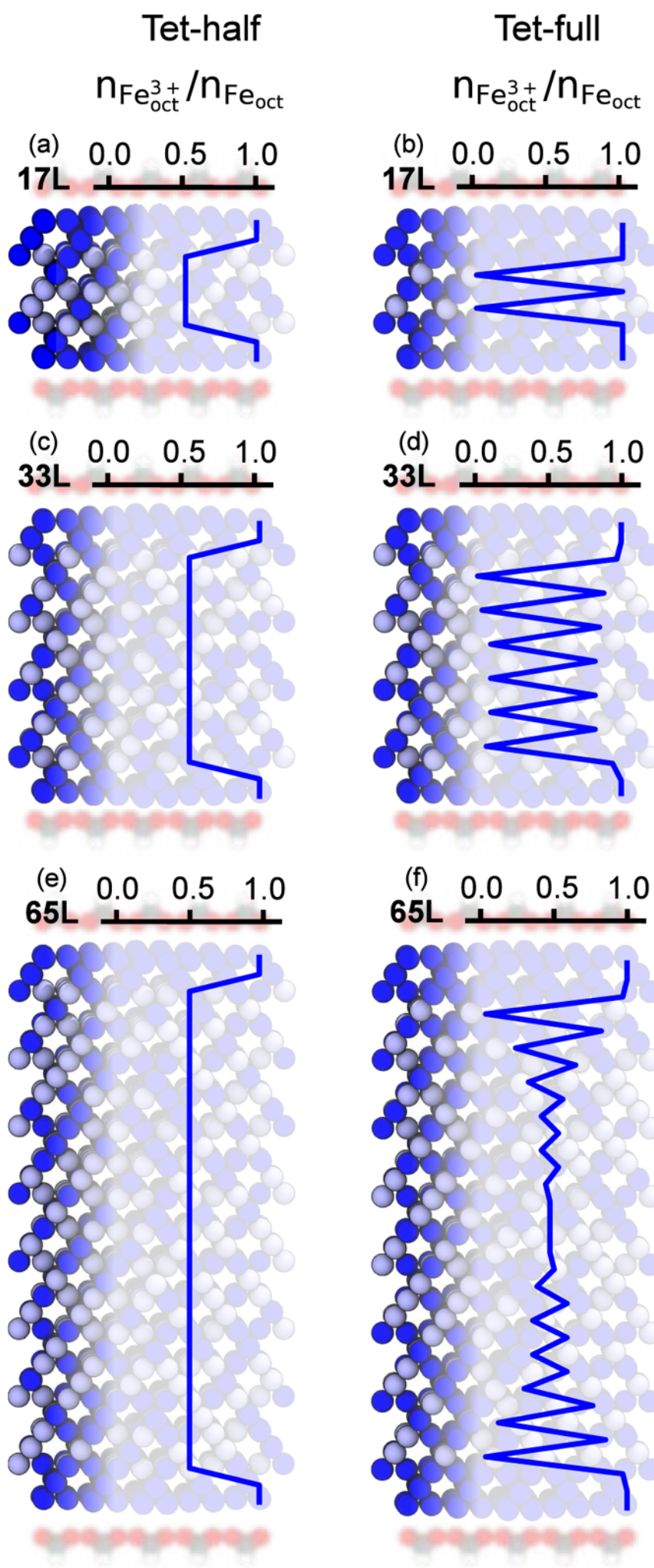

**Figure 3.** Oxidation state distribution of oxidation state minimized formate-magnetite interfaces. The distributions are shown for the formate binding site "tet" and for half and full formate coverage in the left and right columns, respectively. Three slab thicknesses were used: 17L for (a) and (b), 33L for (c) and (d), and 65L for (e) and (f). The number of atomic layers is indicated in the top-left corner of each subfigure. From this perspective half and full coverages look identical. The $Fe^{3+}_{oct}$ ratio within each octahedral layer is denoted by $n_{Fe_{oct}{}^{3+}}/n_{Fe_{oct}}$. $Fe_{tet}$, O (magnetite), $O_H$, and $H_O$ are omitted. Color code: $Fe^{3+}_{oct}$, dark blue; $Fe^{2+}_{oct}$, light blue; $O_C$, red; $H_C$, white; $C_C$, gray.

At 17L and half formate coverage, the first two $Fe_{oct}$ layers at the interface only contain $Fe^{3+}$ oxidation state (cf. Figure 3a). These $Fe_{oct}$ layers at the interface which are dominated by the $Fe^{3+}$ are referred to as "surface layers". The dominance of $Fe^{3+}$ at surface layers is frequently observed, as in bare magnetite[35,39,41] and at magnetite/carboxylic acid interfaces.[24,26] At full coverage, the number of surface layers increases to three (Figure 3b). This pattern continues up to magnetite slabs with 33L: two and three surface layers at half and full coverage, respectively. At full coverage, as there are more negatively charged adsorbates and more $Fe^{3+}$ ions due to our charge neutrality approach (cf. Figure 2c), a greater amount of $Fe^{3+}$ is attracted to the interface, thus increasing the surface layer thickness. The surface layers no longer contain only $Fe^{3+}$ for thicker slabs starting from 33L for full formate coverage. Some $Fe^{2+}$ appear in the third surface layer (cf. Figure 3d).

At half coverage, bulk-like layers with $n_{Fe_{oct}{}^{3+}}/n_{Fe_{oct}} = 0.5$ were observed for all slab thicknesses. In contrast, at full coverage we observe varying degrees of oxidation state layering. The layering becomes less pronounced with increasing slab thickness. For thin slabs, e.g., 17L in Figure 3b, alternating layers exhibiting purely $Fe^{3+}$ or $Fe^{2+}$ were obtained in the simulations. This ideal layering disappears at 33L shown in Figure 3d and a bulk-like region starts to emerge in the center of the slab at 65L given in Figure 3f.

In order to check if systems with nonideal layering or a bulk-like region are energetically more favorable than ideal-layered, we forced thicker slabs to be in a layered configuration and compared resulting energies to those of oxidation state minimized structures in Figure 4a. Minimized structures become increasingly favorable with increasing thickness independent of the binding site.

Subsequently, we investigated the binding site preference, denoted as $\Delta E_{tet\text{-}int}$, of the minimized structures and compared our findings with DFT calculations. At half coverage, we observed a strong "int" preference in Figure 4b which is in good agreement with DFT calculations at 9, 17, and 25L (cf. Tab SI2). During the oxidation state minimization at half coverage, formate molecules diffused from "tet" to "int" binding sites both at 25L and 33L thickness. We refrain from artificially fixing those and decided to exclude those results.

At full coverage, an "int" preference was observed for the thinnest slab with 9L in Figure 4c while DFT suggested "tet". This could be attributed to the fact that our charge neutrality approach at 9L does not leave any $Fe^{2+}$ ions after adsorption. Thicker slabs suggested a slight "tet" preference with 0.003 eV/formate, which agrees reasonably well with our previous empirical force field,[26] that also showed a "tet" preference by 0.02 eV/formate, and with our previous[24] and current DFT calculations, that also showed a "tet" preference by 0.041 eV/formate and 0.047 eV/formate, respectively. The "int" binding seems again to be more favorable for larger slabs at full coverage, where 41L shows an unexpectedly strong preference of "int" binding. Problems with the minimization of such a large system could be a reason for this unexpected behavior. In enforced ideal-layered slabs, "tet" preference was observed regardless of the surface thickness (see the green line in Figure 4c). The thickness of the slab and hence the oxidation state distribution seem to affect to some extent the adsorption behavior of formate, which is an effect previously not considered in many magnetite-adsorbate studies.

Next, we investigated the effects of finite temperatures on binding site preferences. Starting from the minimized structures, we applied Molecular Dynamics (MD) on the structures at 300 K, using the same settings as in our previous study.[39] For each structure, we performed an MD simulation for 10 ns and used the last 5 ns to calculate the binding site preference shown in Figure 4b,c. The results indicate that room temperature has no significant influence on the binding site preferences. It should be noted that the oxidation state distribution was fixed in these MD simulations, since the effects of temperature on the adsorbates, such as changes in adsorption sites, were the main point of interest here.

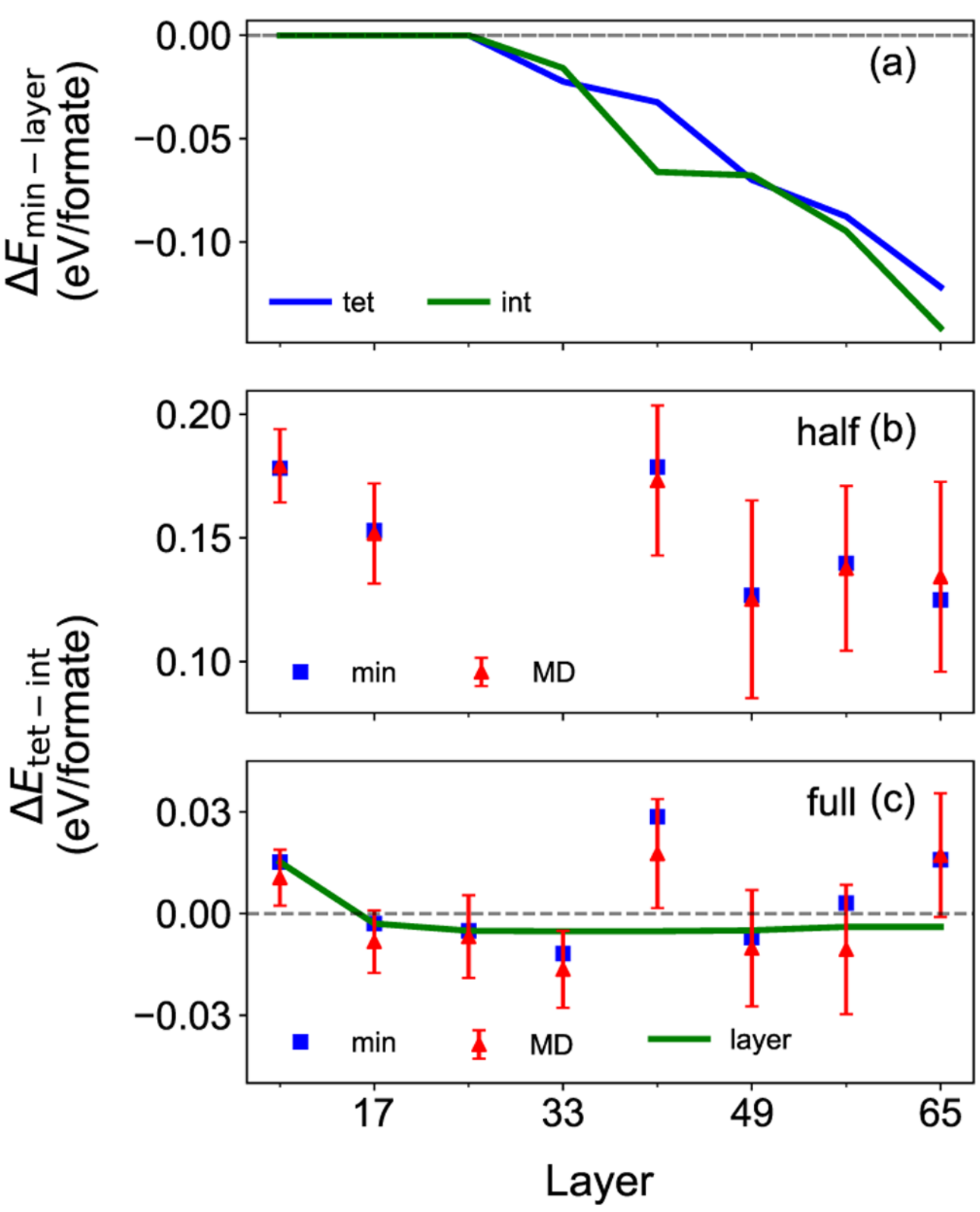

**Figure 4.** (a) Energy difference between oxidation state minimized (min) and layered (layer) structures at full coverage with respect to surface thickness, at "tet" and "int" binding sites. (b) Binding site preference at half coverage for minimized and Molecular Dynamics (MD) resulted structures with respect to surface thickness. Error bars represent the standard deviations of MD. (c) Binding site preference at full coverage of minimized, layered, and Molecular Dynamics (MD) resulted structures with respect to surface thickness.

For oxidation state minimized bare (001) DBT surfaces, the average distance between octahedral layers ($d_{oct} = L_S/(n_{oct} - 1)$) increases with the slab thickness $L_S$, approaching the bulk value.[39] $n_{oct}$ is the number of octahedral Fe layers. For functionalized surfaces, the opposite trend is observed. The average distance $d_{oct}$ is above the bulk value and it decreases with slab thickness toward the bulk value (cf. Figure 5). While at a bare surface relaxations due to surface stresses are causing decreased interlayer distances, formic acid adsorption reduces these stresses and even yields an expansion of the slab. The largest deviation from the bulk value is observed between the first two octahedral Fe layers. This behavior is in very good agreement with our DFT calculations. For the 9L slab, deviations are observed due to limitations of the model, as discussed above. Moreover, it has to be noted that the bare DBT surface is an artificial system since an SCV reconstruction would occur at such surfaces under typical experimental conditions. For the funtionalized DBT surface, the results on the interlayer distances agree well with previous experimental observations.[24] The behaviors for all investigated coverages and adsorption sites are presented in the SI (see Figure SI5).

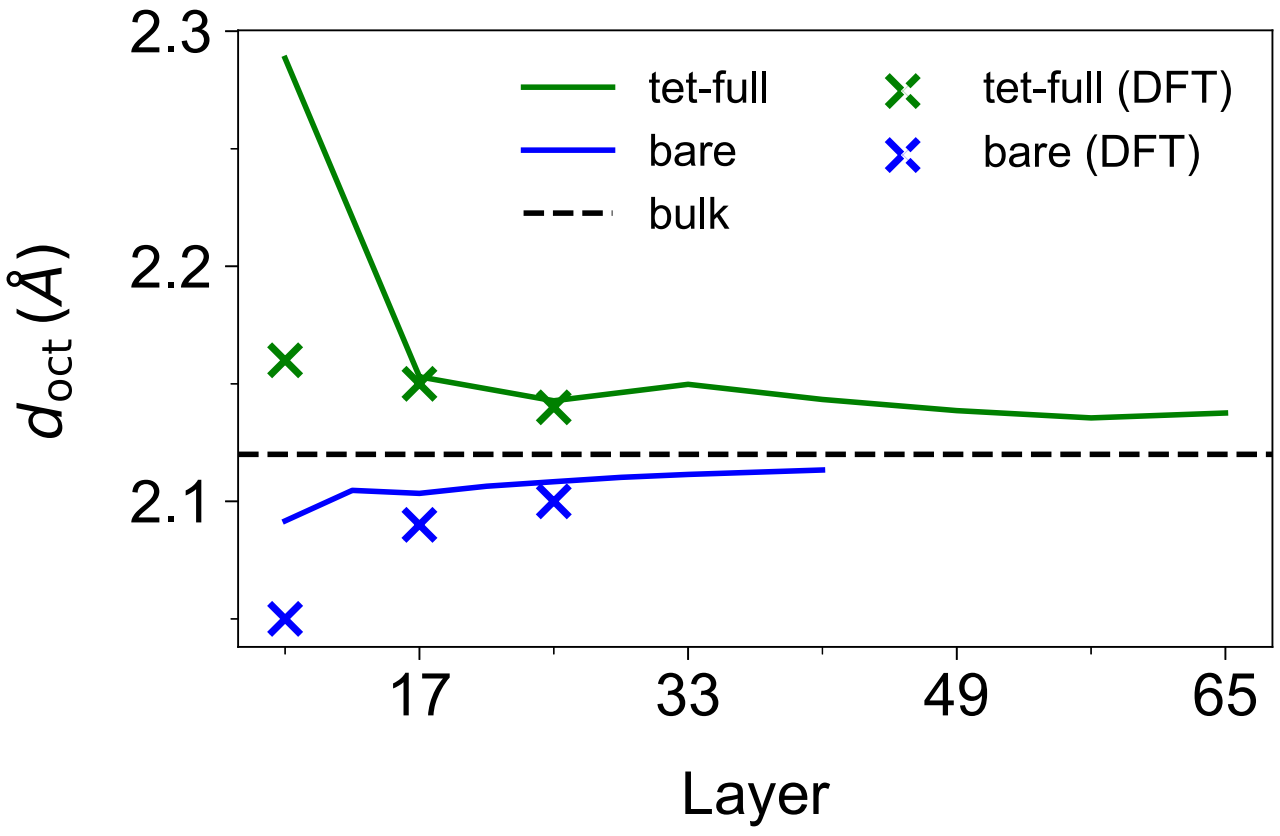

**Figure 5.** The average distance between octahedral layers ($d_{oct}$) as a function of surface thickness. $d_{oct}$ values of oxidation state minimized structures: at full coverage with "tet" binding site, bare surface, and bulk magnetite, with DFT values given for comparison. Additional binding sites and coverages are shown in Figure SI5.

Further, we investigated the bond distances in the magnetite slabs and those with the adsorbates. For the bond length between $Fe_{oct}$ and nearest neighbor oxygens an average value of 2.06 Å was obtained for the central layer of a 25L formate-covered magnetite slab. This underestimates the bulk DFT value by about 1%. Changing the slab thickness, the value showed small modifications from 2.04 Å for 17L to 2.08 Å for 65L. In the surface layer the simulations resulted in distances of 2.03 and 2.14 Å toward surface O and $O_H$, respectively, overestimating DFT and experimental results[24] by about 1.5%. All of those values are in very good agreement with DFT and experiments. The only larger discrepancy was observed for the distance of the surface $Fe_{oct}$ to the nearest oxygen beneath in the subsurface layer. The obtained value of 2.64 Å significantly overestimates the values of 2.19 and 2.14 Å from DFT and experiments,[24] respectively. This value is also overestimated by DFTB and other force field approaches.[28] This deviation occurs because the surface $Fe_{oct}$ in the vicinity of adsorbed formate molecules are lifted too strongly out of the surface plane, indicating an imbalance between the bonds toward the adsorbate and within the slab. Moreover, we investigated the average height of the formate molecules above the magnetite surface, specifically measured between surface $Fe_{oct}$ ions and formate oxygens ($O_C$). Detailed results are shown in the SI (Figure SI6). At full coverage, for both "tet" and "int" binding sites, an average distance of 1.89 Å was observed. For half coverage, the values were slightly increased to 1.91 Å. This is in agreement with our previous studies,[24,26] where the empirical force field yielded 1.88 and 1.89 Å, and DFT gave 2.03 Å. A recent study using MD simulations and DFTB reported distances of 1.92 and 1.99 Å, respectively.[28] Based on SXRD experiments the distance was reported as 2.02(7) Å.[24] It is important to point out that higher-level calculations, DFT[24,26] and DFTB[28] in agreement with experiments,[24] resulted in larger surface−adsorbate distances, which points again to an overestimation of interatomic interactions between adsorbate and the surface in the force fields. In a recent study[28] a reparameterization of the Lennard−Jones (LJ) parameters for the adsorbate oxygen atoms was suggested, which improved those bond lengths but might limit transferability and biocompatibility of the approach. Alternatively, reparameterized LJ parameters of the $Fe^{2+}$ differentiating them from the $Fe^{3+}$ values might be beneficial in general for studying magnetite and its interfaces.

In summary, we have shown that our oxidation state minimization method is capable of modeling magnetite-carboxylic acid interfaces and the results are in good agreement with electronic structure calculations and previous magnetite/carboxylic acid studies.[24,26] Excess charges upon dissociative adsorption, here of formic acid, are compensated by the oxidation of Fe ions. Our MC approach distributes those ions and thus allows the magnetite oxidation state distribution to adapt to interfaces. Particularly, no interface specific parametrization as used in previous approaches[26,28] is required, increasing the applicability and transferability of this approach. As a result, we have determined the number of layers required for an adsorbate-covered surface to have bulk-like layers in the middle of the slab. While at half coverage bulk-like layers are observed even for thin slabs, at full coverage, much thicker slabs are needed. We have shown that the oxidation state distribution depends on the surface thickness and that the binding site preference is affected by the adsorbate coverage. The interlayer spacings for bare and adsorbate-covered surfaces are in very good agreement with the DFT results. Moreover, Fe−O bond lengths in the magnetite bulk, surface, and toward the adsorbates also agree well with DFT, other MD, and experimental results. Only the lifting of surface iron atoms upon adsorption is significantly overestimated, which is a common shortcoming of force field approaches. Updated Lennard−Jones parameters might help to improve this aspect in the future.

The capability to model functionalized magnetite is critical to many applications and thus of high importance. Therefore, the compatibility with biomolecular force fields is essential. By relying on standard force field parameters with only minimal necessary changes for magnetite, a simple charge compensation scheme, and letting the MC approach take care of the adaptation of the oxidation state distribution to the environment, the method stays simple and versatile. The method presented here allows modeling large systems with high accuracy, and thus is suitable, e.g., for studying the mechanical and structural properties of functionalized magnetite based nanocomposites, which will be shown for example in an upcoming work.[42] Moreover, this approach lays the foundation of further hybrid MC/MD simulations, which allow detailed insights into the dynamic behavior of the oxidation states, e.g., at elevated temperatures. In conclusion, the oxidation state minimization method, which is computationally much cheaper than the electronic structure calculation, is a promising tool for modeling magnetite-carboxylic acid interfaces and magnetite-organic adsorbate interfaces in general.

## ASSOCIATED CONTENT

### Supporting Information

The Supporting Information is available free of charge at https://pubs.acs.org/doi/10.1021/acs.jpclett.5c00679.

> Detailed information on the computational settings, electron localization function using density functional theory, and magnetite and carboxylic acid force field parameters (PDF)

## AUTHOR INFORMATION

### Corresponding Author

**Gregor B. Vonbun-Feldbauer** − *Institute for Interface Physics and Engineering and Institute of Advanced Ceramics, Hamburg University of Technology, 21073 Hamburg, Germany;* orcid.org/0000-0002-9327-0450; Email: gregor.feldbauer@tuhh.de

### Authors

**Emre Gürsoy** − *Institute for Interface Physics and Engineering, Hamburg University of Technology, 21073 Hamburg, Germany;* orcid.org/0009-0008-6042-1364

**Robert H. Meißner** − *Institute for Interface Physics and Engineering, Hamburg University of Technology, 21073 Hamburg, Germany; Institute of Surface Science, Helmholtz-Zentrum Hereon, 21502 Geesthacht, Germany;* orcid.org/0000-0003-1926-114X

Complete contact information is available at: https://pubs.acs.org/10.1021/acs.jpclett.5c00679

### Notes

The authors declare no competing financial interest.

## ACKNOWLEDGMENTS

This research was funded by the Deutsche Forschungsgemeinschaft (DFG, German Research Foundation)−SFB 986−192346071.